\documentclass[aps,prb,groupedaddress,twocolumn]{revtex4-2}

\usepackage{graphicx}
\usepackage{color}
\usepackage{nicefrac}
\usepackage{amsmath}
\usepackage{amssymb}
\usepackage{hyperref}
\usepackage{siunitx}
\AtBeginDocument{\usepackage{booktabs}}

\newcommand{\avg}[1]{\left<#1\right>} 
\renewcommand{\vec}[1]{\ensuremath{\mathbf{#1}}} 
\newcommand{\ket}[1]{\big| #1 \big>} 
\newcommand{\braket}[2]{\big<#1 \vphantom{#2} \big| #2 \vphantom{#1} \big>} 
\newcommand{\matrixel}[3]{\big< #1 \vphantom{#2#3} \big| #2 \big| #3 \vphantom{#1#2} \big>} 
\newcommand{\lr}[1]{\left(#1\right)}

\begin{document}

\title{Noncollinear phase of the antiferromagnetic sawtooth chain}

\author{Roman Rausch}
\author{Christoph Karrasch}

\affiliation{Technische Universit\"at Braunschweig, Institut f\"ur Mathematische Physik, Mendelssohnstraße 3,
38106 Braunschweig, Germany}

\begin{abstract}
The antiferromagnetic sawtooth chain is a prototypical example of a frustrated spin system with vertex-sharing triangles, giving rise to complex quantum states.
Depending on the interaction parameters, this system has three phases, of which the gapless non-collinear phase (for strongly coupled basal spins and loosely attached apical spins) has received little theoretical attention so far.

In this work, we comprehensively investigate the properties of the non-collinear phase using large-scale tensor network computations which exploit the full SU(2) symmetry of the underlying Heisenberg model. We study the ground state both for finite systems using the density-matrix renormalization group (DMRG) as well as for infinite chains via the variational uniform matrix-product state (VUMPS) formalism. Finite temperatures and correlation functions are tackled via imaginary- or real time evolutions, which we implement using the time-dependent variational principle (TDVP).

We find that the non-collinear phase is characterized by a low-momentum peak and a diffuse tail for the apex-apex correlations. Deep into the phase, the pattern sharpens into a peak indicating a 90-degree spiral. The apical spins are soft and highly susceptible to external perturbations; they give rise to a large number of gapless magnetic states that are polarized by weak fields and cause a long low-temperature tail in the specific heat. The dynamic spin-structure factor exhibits additive contributions from a two-spinon continuum (excitations of the basal chain) and a gapless peak at $k=\pi/2$ (excitations of the apical spins). Small temperatures excite the gapless states and smear the spectral weight of the $k=\pi/2$ peak out into a homogeneous flat-band structure. Our results are relevant, e.g., for the material atacamite Cu$_2$Cl(OH)$_3$ in high magnetic fields.
\end{abstract}

\maketitle

\section{\label{sec:intro}Introduction}

\begin{figure}[b]
\begin{center}
\includegraphics[width=\columnwidth]{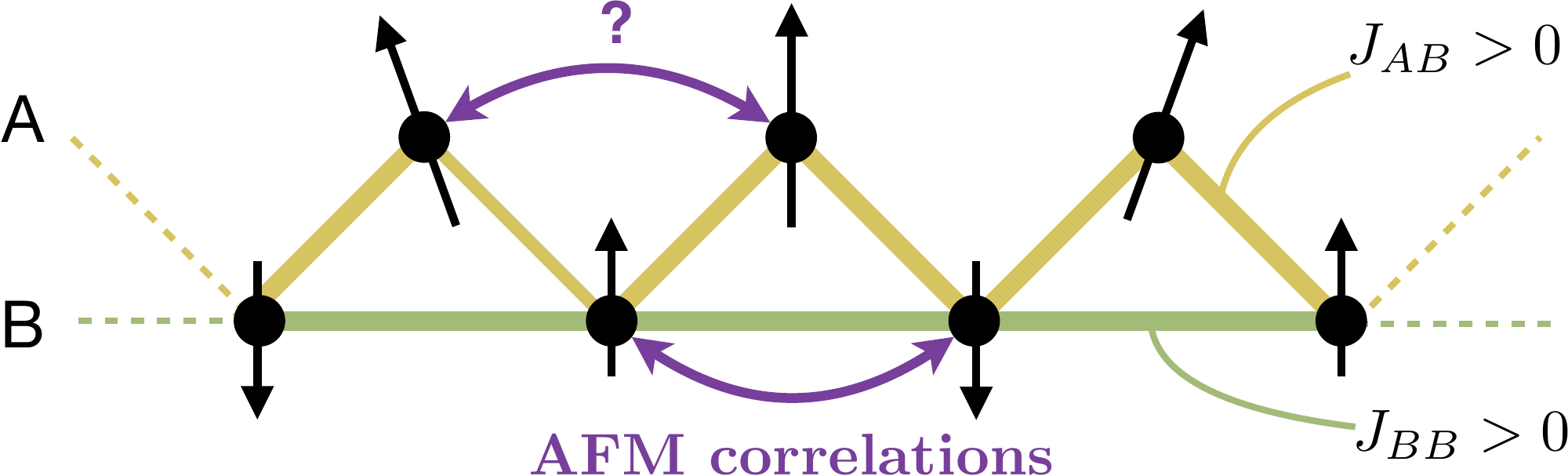}
\caption{\label{fig:sketchSawtooth}
Sketch of the geometry of the sawtooth chain. The two sites A (B) in the unit cell are the apical (basal) spins. The base-base (apex-base) coupling is denoted by $J_{BB}$ ($J_{AB}$).
The apical spins are not directly coupled and develop nontrivial correlations.
}
\end{center}
\end{figure}

Geometrically frustrated magnetic systems are of great interest as a route to a multitude of interesting quantum states and phenomena. A small prototypical example are three spins on a triangle: the classical ground state has a 120-degree alignment; in the quantum case the energy is minimized by degenerate choices of a singlet bond and a free spin.
From this we already learn that geometrical frustration tends to favour degeneracy, pair-singlets and non-collinear states. The interplay of these three factors can be studied in a many-body setting by coupling triangle plaquettes with shared vertices~\cite{Monti_1991}. One of the simplest systems of this kind is the ``sawtooth chain'', a 1D systems with two sites ($A,B$) per unit cell described by the Hamiltonian
\begin{equation}
H = J_{AB} \sum_{i} \lr{\vec{S}^A_i\cdot\vec{S}^B_i + \vec{S}^A_i\cdot\vec{S}^B_{i+1}} + J_{BB} \sum_{i} \vec{S}^B_i\cdot\vec{S}^B_{i+1},
\label{eq:H}
\end{equation}
where $\vec{S}^{A}$ ($\vec{S}^{B}$) is the vector of apical (basal) spin operators; $J_{BB}$ and $J_{AB}$ denote the exchange interaction within the basal chain and between the basal and apical spins, respectively (see Fig.~\ref{fig:sketchSawtooth}). This geometry is realized in a number of solids and molecules~\cite{RuizPerez_2000,Baniodeh_2018,Heinze_2021,Heinze2018}.

The sawtooth chain family already exhibits the promised multitude of phenomena: If $J_{AB}$ is ferromagnetic (``FM-AFM chain''~\cite{Tonegawa_2004,Inagaki_2005,Kaburagi_2005,Yamaguchi_2020}), one finds a gapless phase for $J_{BB}\geq 0.5\big|J_{AB}\big|$ with large-wavelength incommensurate correlations~\cite{Rausch2023}. At $J_{BB}=0.5\big|J_{AB}\big|$, the ground state has macroscopic ground-state degeneracy for all values of the total spin~\cite{Rausch2023,Reichert2023}.

\begin{figure}
\begin{center}
\includegraphics[width=\columnwidth]{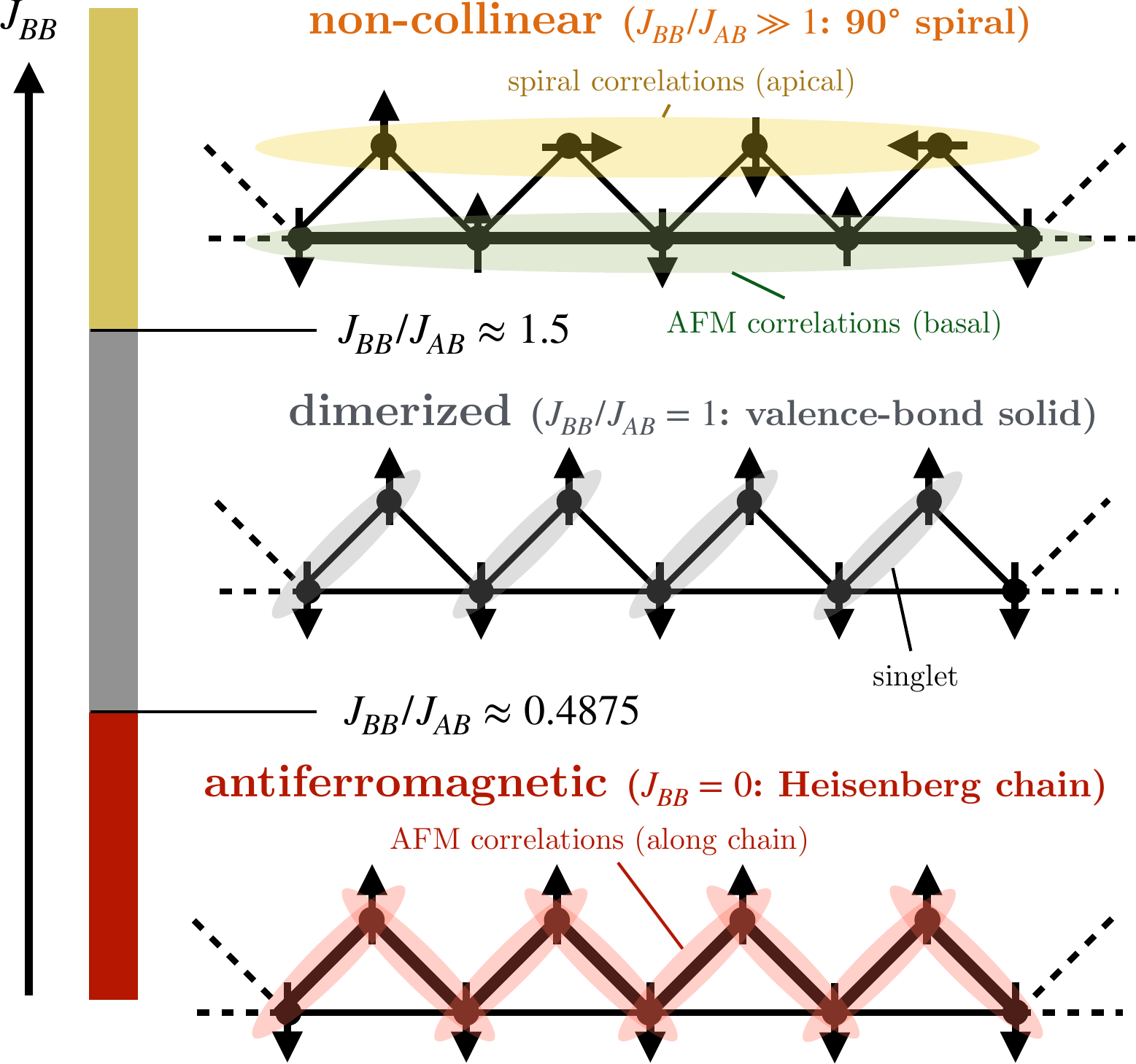}
\caption{\label{fig:PD}
Phase diagram of the AFM-AFM sawtooth chain when $J_{BB}$ is increased from bottom to top: (i) gapless antiferromagnetic phase, adiabatically connected to the AFM Heisenberg chain, (ii) gapped dimerized phase, which for $J_{AB}=J_{BB}$ is an exact valence-bond solid (VBS) of pair-singlets, and (iii) gapless non-collinear phase. The latter is the focus of this work; it is characterized by a commensurate 90-degree spiral of the apical spins for large $J_{BB}$. For estimates of the critical couplings, see Refs.~\cite{Blundell_2003,Jiang_2015A} and Tab.~\ref{tab:Jcrit}.
}
\end{center}
\end{figure}

Here, we focus on the case that both couplings are antiferromagnetic (``AFM-AFM chain''). For this case, the phase diagram is displayed in Fig.~\ref{fig:PD}. For weak $J_{BB}$, the system is in a phase that is adiabatically connected to the simple Heisenberg chain. With increasing $J_{BB}$, it undergoes a transition to a gapped dimerized phase. At the special point $J_{AB}=J_{BB}$, the ground state is a lowly entangled valence-bond solid and the ground-state wavefunction can be given analytically. It is characterized by topological ``kink'' excitations~\cite{Nakamura_Kubo_1996,Sen_1996,Blundell_2003}, whereby broken singlets mediate between left- and right-oriented valence bonds. This phase also exhibits localized-magnon states, which are related to flat bands~\cite{Derzhko_2020}, a huge magnetization jump to half-saturation and a corresponding wide magnetization plateau~\cite{Richter_2005,Richter_2004,Schulenburg_2002,Zhitomirsky_2004}.

If $J_{BB}$ is increased even further, the gap closes around $\left(J_{BB}/J_{AB}\right)_c\approx1.5$~\cite{Blundell_2003} (cf. Fig~\ref{fig:PD}). The emerging gapless phase has received little theoretical attention up to now. Ref.~\cite{Jiang_2015A} mainly used the approximative coupled-cluster method (CCM) along with some exact diagonalization and density-matrix renormalization group (DMRG) numerics for chains of $L\leq120$ sites with open boundary conditions (OBC). It was found that the ground state exhibits non-collinear correlations; it was characterized as ``quasi-canted''. The coupling range considered in \cite{Jiang_2015A} is up to about $J_{BB}/J_{AB}\sim 1.55-1.75$, and no dynamic or thermodynamic properties were computed.

Additional motivation to study the non-collinear phase in more detail comes from recent experiments and first-principle calculations for atacamite~\cite{Heinze2018,Heinze_2021}, which suggest that this material forms nearly decoupled sawtooth chains in large external fields with $J_{BB}=336~\text{K}$, $J_{AB}=102~\text{K}$. This corresponds to $J_{AB}/J_{BB}=0.30$ or $J_{BB}/J_{AB}=3.29$, a value deep in the non-collinear phase. To the best of our knowledge, the only preceding work that has dealt with the systems's properties deep in that phase is Ref.~\cite{Hutak2023}, which used the approximative rotation-invariant Green's function method (RGM) supplemented by exact diagonalization for up to $L=36$ sites (18 unit cells) to study the static thermodynamic properties, as well as the dynamic spin structure factor at finite temperature for the $J$ ratio of atacamite. The main find is a double-peak structure in the specific heat.

In this paper, we undertake a comprehensive study of the gapless non-collinear phase of the model~\eqref{eq:H}. We compute ground-state properties, the magnetization curve, the static spin structure factor (SSF), the specific heat and entropy at finite fields, as well as the dynamic spin-structure factor for $T\geq0$. We briefly discuss the relevance of some of our results for atacamite. Our main aim is to complete the theoretical understanding of the phase diagram of the model~\eqref{eq:H}.

Our results can be summarized as follows. We find that the previously reported ``quasi-canting'' is reflected in the static SSF for the apical spins as a low-momentum peak and a diffuse tail. However, for even stronger coupling the pattern sharpens into one peak with $k=\pi/2$. Thus, the system crosses over to a commensurate 90-degree spiral. In general, the physics of the system is marked by the softness of the apical spins, which are only coupled indirectly via the base and have no direct bonds linking them. They are hence strongly susceptible to even small external fields or temperatures and undergo complex quasi-orderings.

\begin{table}
\centering
\begin{tabular}{ll}
\toprule
authors & $J_c$\\
\midrule
Blundell et al.~\cite{Blundell_2003} & 1.53(1)\\
Jiang et al.~\cite{Jiang_2015A} & 1.42\\
this work & 1.50(7)\\
\bottomrule
\end{tabular}
\caption{
\label{tab:Jcrit} Estimates of the critical ratio $J_c=(J_{BB}/J_{AB})_c$ for the transition between the dimerized phase and the gapless phase.
}
\end{table}

\section{\label{sec:tech}Technical details}

Numerical methods based on matrix-product states (MPS) have become the de-facto standard method to analyze 1D systems~\cite{White_1992,McCulloch_2007,Schollwoeck_2011,Zauner-Stauber_2018}. They exploit the favourable entanglement properties in 1D to obtain highly accurate results controlled by the parameter of the ``bond dimension'' $\chi$. The method is variational and $\chi$ reflects the number of the variational parameters in the wavefunction.

We analyze both large finite and infinite chains using MPS-based methods.

\subsection{Finite systems, ground state}
For finite systems, we employ the DMRG and fully exploit the SU(2) symmetry of the Hamiltonian~\eqref{eq:H}~\cite{McCulloch_2007}. The resulting gain in the bond dimension allows us to study systems with periodic boundary conditions (PBC) of up to $L=2\times100$ sites ($100$ unit cells). The PBC are implemented by representing the Hamiltonian as a matrix-product operator (MPO) with longer-ranged interactions. We use the two-site DMRG algorithm to grow the bond dimension before switching to the cheaper single-site algorithm with perturbations~\cite{Hubig_2015}.

To gauge the error, we compute the energy variance per site:
\begin{equation}
\text{var}/L = \big|\avg{H^2}-\avg{H}^2\big|/L.
\label{eq:var}
\end{equation}
Setting the largest energy scale to 1, we find that $\chi_{\text{SU(2)}}=2000$ (corresponding to $\chi_{\text{eff}}\sim8000$ without SU(2) symmetry) is enough to achieve $\text{var}/L \leq 10^{-5}$.

The finite-system approach is useful to directly target the lowest state in each sector of the total spin $S_{\text{tot}}$, which is a conserved quantum number:
\begin{equation}
\left\langle\vec{S}_{\text{tot}}^2\right\rangle = \sum_{\alpha,\gamma=A,B}\sum_{ij} \left\langle\vec{S}^{\alpha}_i\cdot\vec{S}^{\gamma}_j\right\rangle = S_{\text{tot}}\lr{S_{\text{tot}}+1}.
\end{equation}
Because of the AFM couplings, we expect the ground state without an external field to be in the $S_{\text{tot}}=0$ sector (we have verified this exemplary). Our SU(2)-symmetric algorithm directly yields all $2S_\text{tot}+1$ members of the ground state manifold (labelled by $M_\text{tot}$). For all data shown in this paper, it is irrelevant which member of the ground state manifold is used when computing ground-state expectation values $\langle\cdot\rangle$.

The Hamiltonian in the presence of a magnetic field $B$ reads:
\begin{equation}
H_B = H - B \sum_i \left(S^{z,A}_i+S^{z,B}_i\right).
\label{eq:HB}
\end{equation}
Since the eigenstates of $H$ are also the eigenstates of $H_B$, we can treat magnetic fields for finite systems without explicitly breaking the SU(2) symmetry by considering the energies
\begin{equation}
E_B(S_{\text{tot}},M_{\text{tot}}) = E_0(S_{\text{tot}}) - BM_{\text{tot}},
\label{eq:EB}
\end{equation}
where $E_0(S_{\text{tot}})$ are the lowest energies of $H$ in each sector of $S_{\text{tot}}$ (independent of $M_{\text{tot}}$). Minimizing with respect to $M_{\text{tot}}$ yields $M_{\text{tot}}=S_{\text{tot}}$, which removes one parameter. The final minimization has to be done numerically for each fixed value of $B$.

\subsection{Infinite systems, ground state}
For infinite systems, we use the MPS-based variational uniform matrix-product states (VUMPS) formalism~\cite{Zauner-Stauber_2018}. By exploiting the SU(2) symmetry, it allows us to study the infinite chain for $S_{\text{tot}}=0$ where we can similarly reach $\chi_{\text{eff}}\sim10,000$. Our numerical unit cell encompasses two physical unit cells (4 sites) in order to allow for dimerization.

In the presence of an external magnetic field (Eq.~\ref{eq:HB}), we switch off the symmetry altogether and directly compute the resulting magnetization~\eqref{eq:Mtot} with an explicit $B$-term present, but the bond dimension is limited by $\chi\sim500$. The cross-diagnostic combination of all the approaches allows us to form a full picture of the system.

\subsection{Finite temperatures}
For finite temperatures, we use the standard approach of doubling the degrees of freedom and purifying the density matrix~\cite{FeiguinWhite2005}. The state at $\beta=1/T=0$ can be exactly prepared and propagated in imaginary time: $\ket{\beta}=\exp\lr{-\beta H/2}\ket{\beta=0}$. We employ the two-site time-dependent variational principle (TDVP) algorithm~\cite{Haegeman2016} with open boundary conditions (OBC). The discarded weight is kept fixed, and the bond dimension increases dynamically. We exploit the SU(2) spin-rotational symmetry for $B=0$ and the U(1) symmetry for $B\neq0$. The specific heat per site is obtained from
\begin{equation}
\frac{C\lr{T}}{L} = \frac{1}{L} \beta^2 \big[ \avg{H_B^2}_{\beta} - \avg{H_B}_{\beta}^2 \big],
\label{eq:c}
\end{equation}
where $E\lr{\beta} = \avg{H_B}_{\beta} = Z^{-1}_{\beta} \matrixel{\beta}{H_B}{\beta}$ is the internal energy and $Z_{\beta} = \braket{\beta}{\beta} = \matrixel{\beta=0}{e^{-\beta H_B}}{\beta=0}$ is the partition function.
We also compute the entropy density via
\begin{equation}
\frac{S(\beta)}{L} = \frac{\ln Z_{\beta}}{L} + \beta\frac{E\lr{\beta}}{L}.
\label{eq:s}
\end{equation}

\subsection{Correlation functions at zero and finite temperature}
We compute the dynamic properties of the system at $T=0$ and at finite temperatures. The dynamic spin structure factor is related to neutron-scattering experiments and its corresponding $T=0$ retarded Green's function is given by
\begin{equation}
G^{\alpha\gamma}\lr{t,\big|i-j\big|} = -i\theta\lr{t} \big\langle \vec{S}^{\alpha}_i\lr{t}\cdot\vec{S}^{\gamma}_j\big\rangle,
\end{equation}
with $\alpha,\beta=A,B$. Its Fourier transform reads:
\begin{equation}
G^{\alpha\gamma}\lr{\omega,k} = \sum_{d} e^{ikd} \int_{-\infty}^{\infty} dt~ e^{i\omega t} G^{\alpha\gamma}(t,d),
\end{equation}
and we are interested in the trace of the spectral function
\begin{equation}
\text{Tr}~S\lr{\omega,k} = -\frac{1}{\pi} \sum_{\alpha=A,B} \text{Im}~G^{\alpha\alpha}\lr{\omega,k}.
\label{eq:Strace}
\end{equation}
We compute $\text{Tr}~S\lr{\omega,k}$ using infinite boundary conditions~\cite{Phien2012}. From the VUMPS ground state, we assemble a finite section of size $L=2\times32$ on which the local excitation is allowed to propagate until $t_{\text{max}}=24 J_{BB}^{-1}$ (setting $\hbar=1$).

The above formulas can be generalized to finite temperatures in a standard manner~\cite{FeiguinWhite2005,Karrasch2013,Barthel2016,Rausch2021}:
\begin{equation}
G^{\alpha\gamma}\lr{t,\big|i-j\big|} = -i\theta\lr{t} Z\lr{\beta}^{-1} \matrixel{\beta}{ \vec{S}^{\alpha}_i\lr{t}\cdot\vec{S}^{\gamma}_j }{\beta}.
\end{equation}
In the finite-temperature case, we use a finite system (OBC), $L=2\times32$ and a maximal propagation time of $t_{\text{max}}=16 J_{BB}^{-1}$.

Both at $T=0$ and $T>0$, the system size is chosen large enough such that the local propagation does not reach the boundaries. This ensures that the results are free from finite-size errors and unphysical reflections, and are only limited by $t_{\text{max}}$. The real time evolution is computed using the 2-site TDVP, and SU(2) symmetries are exploited. The discarded weight is kept fixed, and the bond dimension increases dynamically.

\begin{figure}
\begin{center}
\includegraphics[width=\columnwidth]{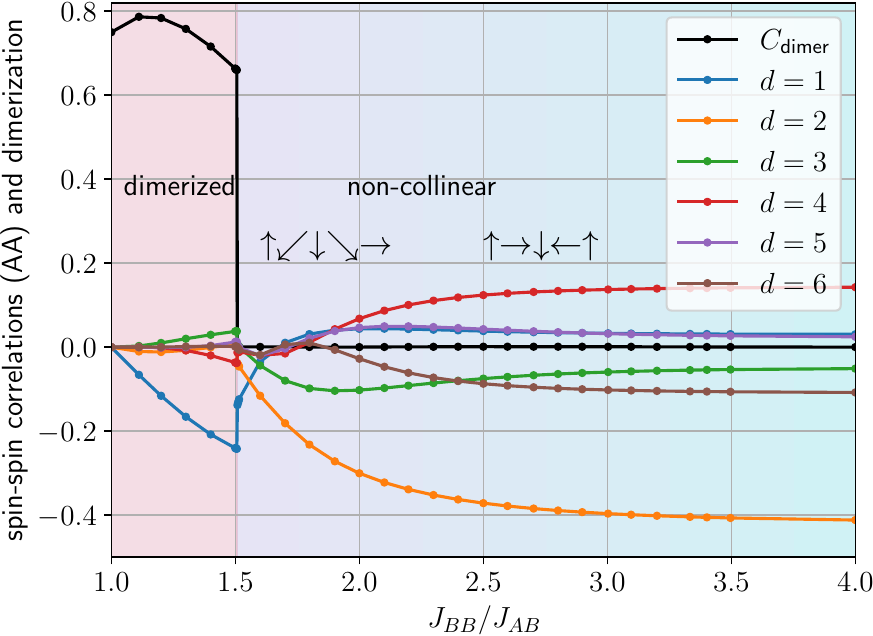}
\caption{\label{fig:corr}
Dimerization order parameter Eq.~\eqref{eq:Cdimer} and apex-apex spin correlations $\avg{\vec{S}^A_i\cdot\vec{S}^A_{i+d}}$ in the ground state calculated using VUMPS for the infinite system.
}
\end{center}
\end{figure}

\section{\label{sec:PT}Transition point from the dimerized phase}

We first determine the critical coupling strength for the phase transition between the gapped dimerized phase and the gapless non-collinear phase (see Fig.~\ref{fig:PD}), which was previously estimated to be $(J_{BB}/J_{AB})_c=1.53(1)$ by extrapolation from finite systems~\cite{Blundell_2003} and $(J_{BB}/J_{AB})_c\approx1.42$ by using the CCM approximation~\cite{Jiang_2015A}.

The dimerized phase breaks spatial symmetry and has an order parameter given by
\begin{equation}
C_{\text{dimer}}= \big| \avg{\vec{S}^A_i\cdot\vec{S}^B_i}-\avg{\vec{S}^A_i\cdot\vec{S}^B_{i+1}} \big|.
\label{eq:Cdimer}
\end{equation}
In such a case, MPS-based methods tend to converge to a particular symmetry-broken minimum (with a lowly entangled state), rather than to a superposition of the states from the two minima. Thus, the order parameter can be directly calculated (note that dot products $\langle {\bf S}_i^\alpha\cdot {\bf S}_j^\gamma\rangle$ do not depend on $M_\text{tot}$). For the infinite system, we find $C_{\text{dimer}}\approx0.66$ for $J_{AB}/J_{BB}=0.664$ and a jump down to $C_{\text{dimer}}\sim10^{-4}$ for $J_{AB}/J_{BB}=0.663$ (see Fig.~\ref{fig:corr}), indicating a first-order phase transition. The transition point is thus estimated as $(J_{AB}/J_{BB})_c=0.663(5)$ or $(J_{BB}/J_{AB})_c=1.50(7)$.

We have corroborated this by studying the correlation length $\xi(\chi)$ obtained from the MPS transfer matrix~\cite{McCulloch2008} as a function of the bond dimension $\chi$. The value of $\xi(\chi)$ should quickly saturate in a gapped phase with a finite correlation length, which can be encoded accurately by an MPS. We find that $J_{AB}/J_{BB}\geq0.664$ is clearly gapped, while $J_{AB}/J_{BB}=0.663$ becomes gapless (data not shown).

The various estimates for $(J_{AB}/J_{BB})_c$ are summarized in Tab.~\ref{tab:Jcrit}.

\begin{figure}
\begin{center}
\includegraphics[width=\columnwidth]{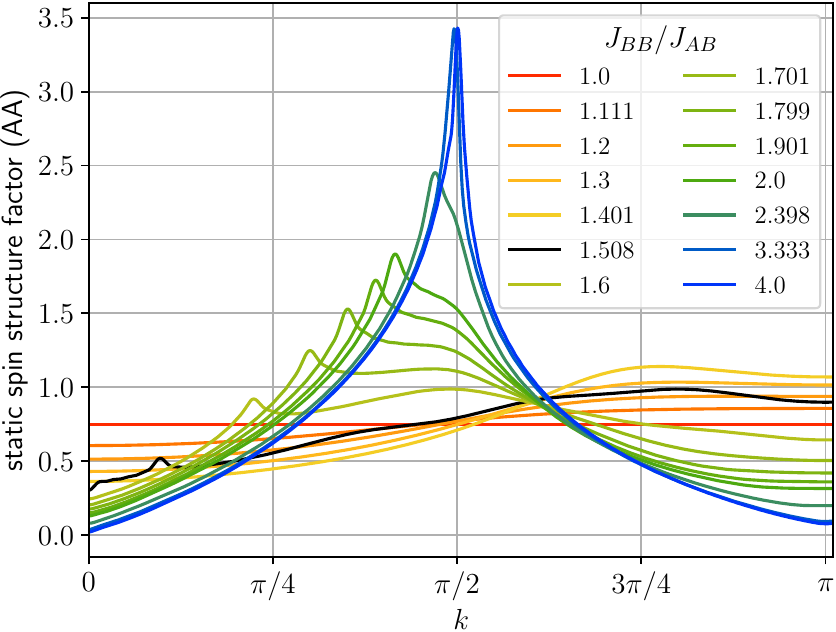}
\caption{\label{fig:SSF}
Static spin structure factor Eq.~(\ref{eq:SSFcosine}) for the apex-apex correlations in the ground state computed using VUMPS for the infinite system. The black line indicates the first value in the non-collinear phase.
}
\end{center}
\end{figure}

\section{\label{sec:gs}Nature of the non-collinear ground state}

For a strong $J_{BB}$, the basal spins always display antiferromagnetic correlations. Understanding the nature of the ground state is thus equivalent to understanding the apical spin correlations (see Fig.~\ref{fig:sketchSawtooth}). They cannot be trivially deduced because there are no direct bonds that connect the apical spins alone.

Figure~\ref{fig:corr} also shows the apical spin correlations $\avg{\vec{S}^A_i\cdot\vec{S}^A_{i+d}}$ up to a distance of 6 unit cells for the infinite system (note that this quantity is independent of $i$). When the non-collinear phase is entered, one observes negative correlations for the nearest neighbours that decrease in magnitude. We can picturize this alignment as $\uparrow\swarrow\downarrow\searrow\rightarrow$, where the first spin is AFM-correlated to a number of its nearest neighbours. In this sense, the previous description as quasi-canted~\cite{Jiang_2015A} seems appropriate for the region close to the transition point.
However, when $J_{BB}$ is increased, this complex regime crosses over to a simpler pattern of larger absolute values and signs $-+-$ for $d=2,4,6$, and smaller absolute values and signs $+-+$ for $d=1,3,5$. This indicates 90-degree correlations that can be visualized as $\uparrow\rightarrow\downarrow\leftarrow\uparrow$.

The corresponding spin-structure factor~\cite{Rausch2023}
\begin{equation}
\text{SSF}(k) = \avg{\mathbf{S}^A_{j_0} \cdot \mathbf{S}^{A}_{j_0}} + 2\sum_{d=1}^{\infty} \cos\lr{kd} \avg{\mathbf{S}^{A}_{j_0} \cdot \mathbf{S}^{A}_{j_0+d}}
\label{eq:SSFcosine}
\end{equation}
is shown in Fig.~\ref{fig:SSF} (the above expression does not depend on $j_0$, and the $d$-sum is performed until convergence is reached). In the dimerized phase, we have a flat pattern due to the short-ranged correlations, with a diffuse peak close to $k=\pi$. In the non-collinear phase, a relatively sharp low-momentum peak appears close to $k=0$.
The resulting pattern is complex and resembles a double-Q structure, which corresponds to quasi-canting in real space.
As $J_{BB}$ is increased, the pattern sharpens into a simpler structure with one dominant peak at $k=\pi/2$, i.e. 90-degree correlations.

\begin{figure}
\begin{center}
\includegraphics[width=\columnwidth]{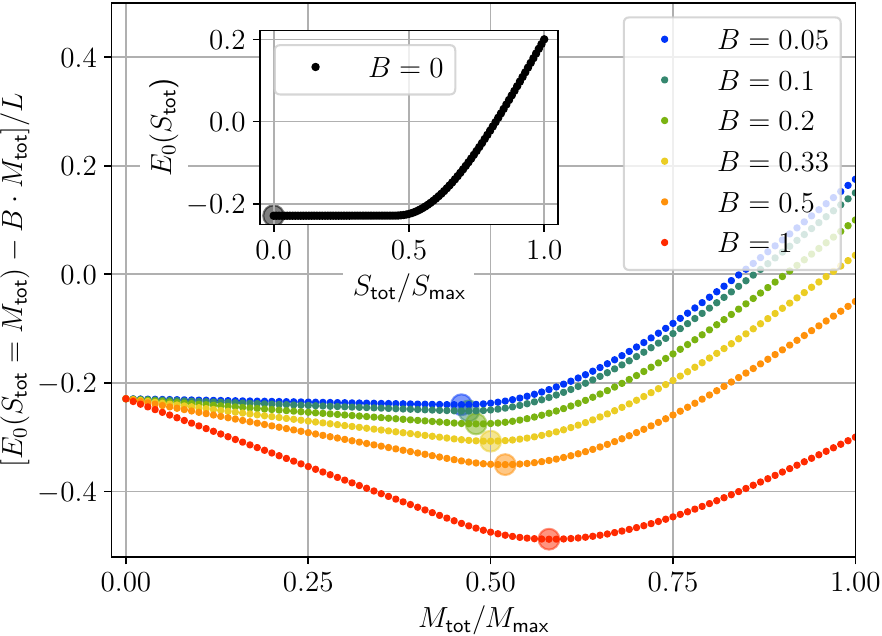}
\caption{\label{fig:EtowerBz}
Ground-state energy~\eqref{eq:EB} per site for $S_{\text{tot}}=M_{\text{tot}}$ as a function of $M_{\text{tot}}$ (normalized to $M_{\text{max}}=S_{\text{max}}=L/2$) in the presence of a B-field for $J_{BB}=1$, $J_{AB}=0.3$.
Inset: The lowest energy in each sector of $S_{\text{tot}}$ at $B=0$ (tower of states).
The semi-transparent circles indicate the position of the energy minimum.
The results were obtained using DMRG with PBC for $L=200$.
}
\end{center}
\end{figure}

\section{\label{sec:tower}Magnetic field properties}

\subsection{The tower of states}

The inset of Fig.~\ref{fig:EtowerBz} shows the ``tower of states'' in the non-collinear phase for $B=0$, i.e., the lowest energies $E_0(S_{\text{tot}})$ in each sector of the total spin. As expected, the curve features a minimum at $S_\text{tot}=0$; moreover, it is basically flat (gapless) up to half-saturation $S_{\text{tot}}/S_{\text{max}}\lesssim0.5$. A non-zero $B$-field thus immediately leads to an energy minimum close to $M_{\text{tot}}/M_{\text{max}}\approx0.5$ (main panel of Fig.~\ref{fig:EtowerBz}).

This observation can be understood as follows: Frustration on the triangles and a relatively small $J_{AB}$ prevents a preferred alignment of the apical spins with respect to the basal ones. They are effectively ``soft'', can be easily flipped, and are unable to resist the field. Thus, the half-saturation minimum is mainly due to the apical spins.
On the other hand, the basal spins are strongly coupled with a preferred AFM alignment. They resist the field more effectively, so that a further shift in $M_{\text{tot}}/M_{\text{max}}$ beyond half-saturation requires much stronger fields.

\begin{figure}[b]
\begin{center}
\includegraphics[width=\columnwidth]{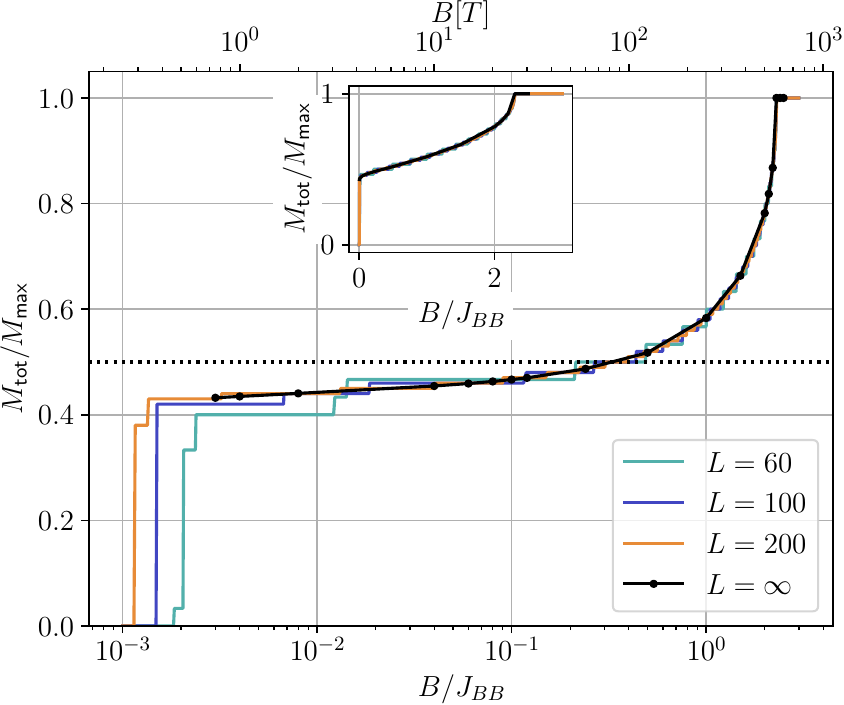}
\caption{\label{fig:mag}
Ground state magnetization as a function of $B$ for $J_{BB}=1$, $J_{AB}=0.3$ (normalized to the saturation value $M_{\text{max}}=L/2$). The dotted line indicates half-saturation.
The inset shows the same without the logarithmic scaling on the x-axis, revealing the absence of magnetization plateaus.
The results were obtained using DMRG for various system sizes $L$ as well as using VUMPS for the infinite systems.
}
\end{center}
\end{figure}

\subsection{Magnetization process}

We now investigate the fact that the apical spins are easily polarized by weak fields in more detail. Figure~\ref{fig:mag} shows the ground state magnetization
\begin{equation}
M_{\text{tot}} = \sum_i \big\langle \big(S^{z,A}_i+S^{z,B}_i\big)\big\rangle
\label{eq:Mtot}
\end{equation}
as a function of $B$. Numerically, we can resolve weak fields of the order of $B/J_{BB}\sim10^{-3}$ that are still enough to immediately polarize the system slightly below half-saturation. There is a systematic shift of the polarization onset with the system size towards smaller $B$. For the infinite system, we find a polarized state with $M/M_{\text{max}}\approx0.43$ down to $B=0.003$.
Similar results led the authors of Ref.~\cite{Heinze_2021} to speculate that the fieldless system might be ordered. We argue that this is not possible for purely AFM interactions in 1D. Our interpretation is rather that large changes in the total spin become gapless (cf. Fig.~\ref{fig:EtowerBz}), so that a significant polarization is achieved by an arbitrarily small field in the thermodynamic limit.

\subsection{Absence of magnetization plateaus}

Figure~\ref{fig:mag} is plotted semi-logarithmically, which creates the illusion of a magnetization plateau. The inset shows a linear plot, indicating that such a plateau is absent in the non-collinear phase. 

We note that the strong couplings for atacamite translate into $1\text{T}$ being equivalent to $\sim0.004J_{BB}$ (using $g=2$); and the maximally reached magnetic field in the experiment is of the order of $60\text{T}$ or $0.24J_{BB}$. Thus, while the apical spins are polarized immediately, the basal coupling is so strong that even a variation in the field of almost two orders of magnitude is unable to significantly affect them. The experimental magnetization curve for atacamite~\cite{Heinze_2021} thus resembles the logarithmic plot in Fig.~\ref{fig:mag} with a pseudo-plateau.

\section{\label{sec:thermodyn}Magnetothermodynamics}

\begin{figure}[t]
\begin{center}
\includegraphics[width=\columnwidth]{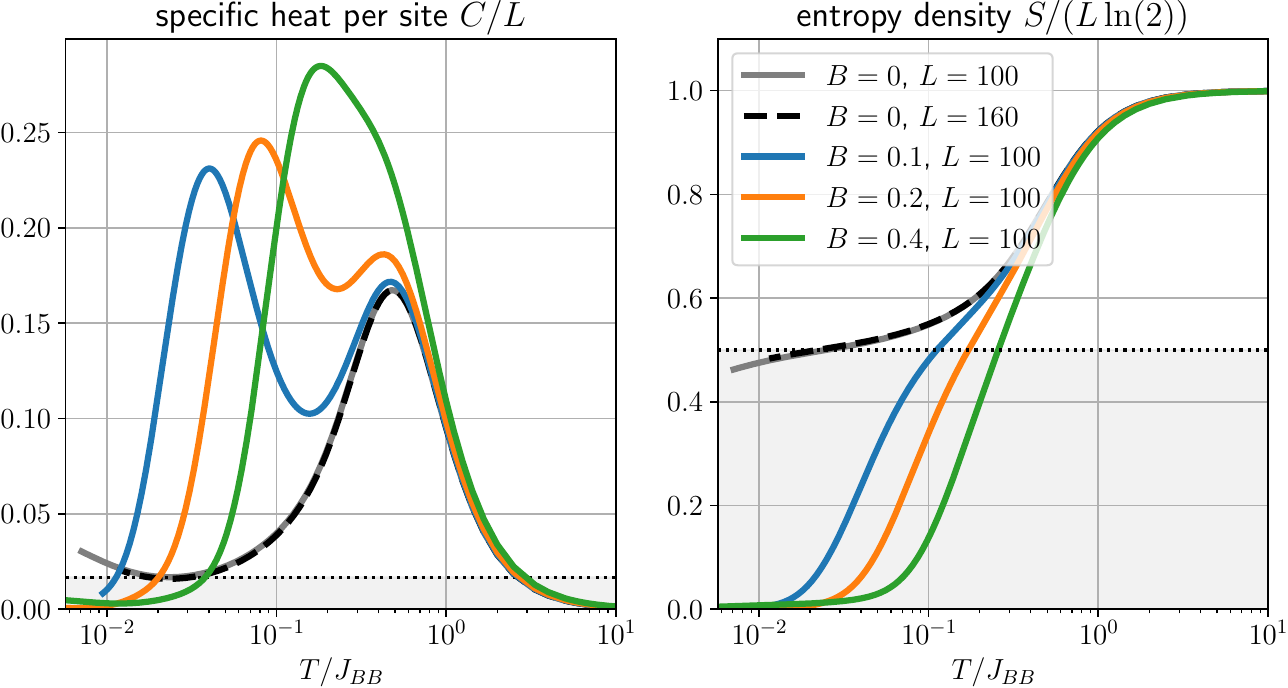}
\caption{\label{fig:c}
Left: Specific heat $C/L$ per site, see Eq.~(\ref{eq:c}), for $J_{BB}=1$, $J_{AB}=0.3$, $L=100,160$ (OBC) and various values of the magnetic field $B$. For $B=0$, the long low-temperature tail is highlighted as the grey shading.
Right: The corresponding entropy density $S/L$ of Eq.~(\ref{eq:s}) normalized by $\ln 2$. For $B=0$, about half of the entropy is unreleased at low temperatures, which is highlighted as the grey-shaded area.
}
\end{center}
\end{figure}

We discuss the thermodynamic properties of the system. The specific heat per site $C/L$ is shown in Fig.~\ref{fig:c}. The peak in $C/L$ around the largest energy scale of $T\sim J_{BB}$ is a typical feature for all spin systems. Particular to our system is a broad tail in $C/L$ for $B=0$ down at least to $T\sim0.02J_{BB}$. Correspondingly, we see that that the system still retains $\sim50\%$ of the entropy at this temperature. We interpret this again as a consequence of the gapless excitations of the apical spins, which result in a high density of states close to the ground state. These states feel $T\sim0.02J_{BB}$ as essentially infinite.
When a finite $B$-field is included, the low-lying magnetic states are raised in energy (see Fig.~\ref{fig:EtowerBz}). This results in a second peak in $C/L$, which is pushed to higher temperatures as $B$ is increased, and eventually merges with the main peak at $T\sim J_{BB}$. Correspondingly, the entropy decreases more quickly, as there is a smaller density of states in the vicinity of the polarized ground state.

The zero-field specific heat was also computed using RGM in Ref.~\cite{Hutak2023}. Interestingly, the method predicts a secondary peak around $T\sim0.02J_{BB}$. Supplementary ED calculations for up to $L=2\times 18$ (PBC) show that $C/L$ is increasing in this temperature region. However, the RGM method is only approximate (it yields a ground-state energy that is higher than the numerically exact VUMPS result), and ED is confounded by finite-size effects that start to set in once the temperature gets to the energy range of the low-lying apical excitations.
In order to address this issue, Fig.~\ref{fig:c} shows the zero-field DMRG result for $L=100,160$ (OBC). $L=100$ has a clear upwards trend around $T\sim10^{-2}J_{BB}$ which remains for $L=160$ and is thus unlikely to be a finite-size effect. This gives additional credence to the existence of a secondary peak at even lower temperatures.

\section{\label{sec:dynamics}Zero-field dynamics}

\begin{figure*}[t]
\begin{center}
\includegraphics[width=0.47\linewidth]{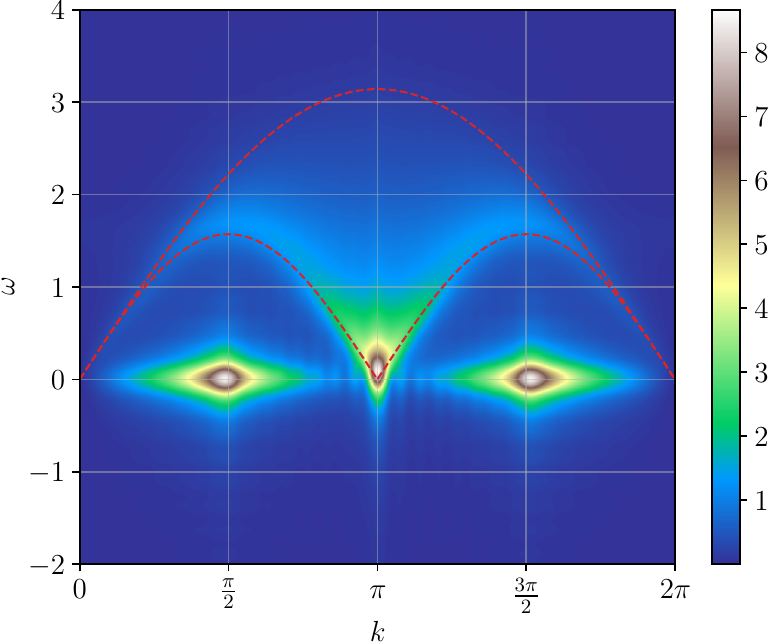}\hspace*{0.03\linewidth}
\includegraphics[width=0.47\linewidth]{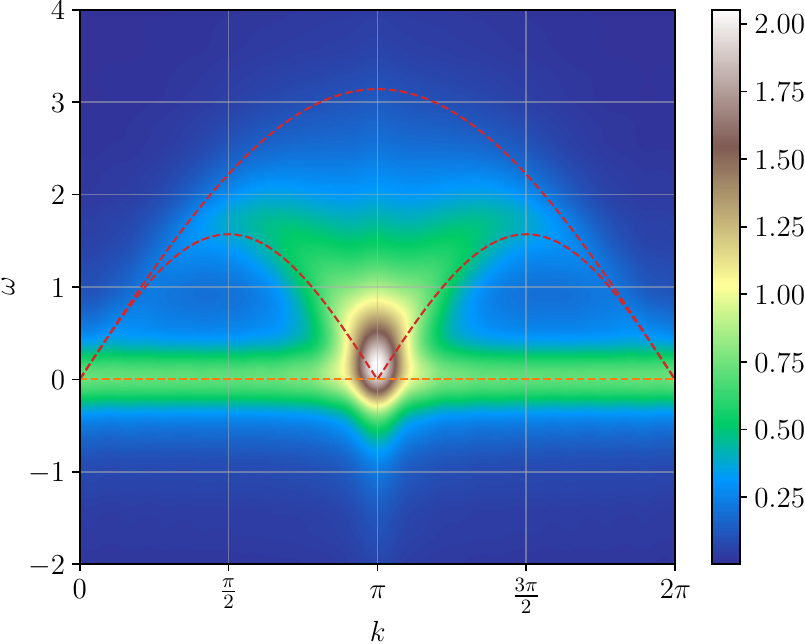}
\caption{\label{fig:kspec}
Left: Trace of the zero-temperature spectral function Eq.~\eqref{eq:Strace} for $J_{BB}=1$, $J_{AB}=0.3$, infinite boundary conditions (propagation on a hetergeneous section of $L=2\times32$ sites). The dotted red line indicates the limits of the two-spinon continuum $\pi J_{BB}/2 \left|\sin\lr{k}\right|$ and $\pi J_{BB}\left|\sin\lr{k/2}\right|$~\cite{desCloizeaux1962} with $J_{BB}=1$. Right: The same as in Fig.~\ref{fig:kspec} but for finite temperatures $T=0.02J_{BB}$ and $L=64$ (OBC).
}
\end{center}
\end{figure*}

The dynamic spin-structure factor in Eq.~\eqref{eq:Strace} for $T=0$ deep in the non-collinear phase is shown in Fig.~\ref{fig:kspec} (left panel). The spectrum is simply an additive contribution of (i) a two-spinon continuum (red dotted line), known as the des Cloizeaux-Pearson modes~\cite{desCloizeaux1962}, which are the excitations of the basal chain, and (ii) a flat band with a strong spectral weight at $k=\pi/2$, which are the excitations of the 90-degree spiral of the apical spins.

The right panel of Figure~\ref{fig:kspec} shows how this changes for a small temperature of $T=0.02J_{BB}$: The spectral weight at $k=\pi/2$ now smears out into a homogeneous flat band. Our interpretation is that even a small temperature is able to excite to gapless continuum of the apical spin states and washes away the pure ground-state contribution of the 90-degree correlations. Thus, the presence of the gapless apical excitations is a pervasive feature in various observables. This also means that the $k=\pi/2$ peak will be difficult to observe experimentally even at very small temperatures.
Finally, we note that RGM calculations for (Fig. 4 in Ref.~\cite{Hutak2023}) also appear to show the incipient $k=\pi/2$ peak for very low temperatures.


\section{\label{sec:summary}Summary}

We have presented a detailed investigation of the gapless non-collinear phase of the AFM-AFM sawtooth chain, where the basal spins are strongly coupled and the apical spins are loosely attached. Few prior studies focused on this regime, and our work completes the understanding of the full phase diagram of the system. Additional experimental motivation comes from atacamite that may show some of the effects discussed here (especially the magnetization curve). However, the role of interchain couplings and potential anisotropies complicate the interpretation of experimental data, which is beyond the scope of this paper.

Our results were obtained using large-scale tensor network numerics which exploit the SU(2) symmetry of the problem. We computed the ground-state properties for finite (DMRG) and infinite (VUMPS) systems; finite temperatures were implemented using ancillas.

The properties of the system are governed by the loose apical spins, which lack direct bonds to optimize the energy. They are thus ``soft'' and easily perturbed, which results in the following effects:
(i) ``quasi-canted'' correlations and a static spin structure factor resembling a double-Q structure close to the phase boundary, but with a diffuse tail instead of a secondary peak, (ii) 90-degree spiral correlations deep in the phase, (iii) gapless excitations with large changes in the total spin, leading to a very easy polarization by an external field and smearing out of the 90-degree peak for small finite temperatures, and (iv) a large density of states manifesting itself in a broad tail of the specific heat and nonvanishing entropy for low temperatures.

The question of a secondary peak in the specific heat for very low temperatures remains difficult. However, a cross-method diagnosis using DMRG for large systems (OBC), as well as preceding work using ED for small systems (PBC) and RGM~\cite{Hutak2023} points to its existence.

Our study adds to the list of diverse phenomena in the sawtooth chain family, further illustrating the complex effects of frustration. The system demonstrates how easily polarizable spins appear despite isotropic and antiferromagnetic interactions, solely via the frustrated geometry.

\begin{acknowledgements}
We thank S. S\"ullow and O. Derzhko for inspiring discussions. C.K. and R.R. acknowledge support by the Deutsche Forschungsgemeinschaft (DFG, German Research Foundation) under Germany's Excellence Strategy EXC-2123 QuantumFrontiers 390837967.
\end{acknowledgements}

\bibliography{sawtooth.bib}

\end{document}